\documentclass[twocolumn,preprintnumbers,amsmath,amssymb]{revtex4}

\usepackage{graphicx}% Include figure files
\usepackage{dcolumn}% Align table columns on decimal point
\usepackage{bm}% bold math

\begin{document}

%\preprint{APS/123-QED}

\title{Giant electrocaloric effect in the thin film relaxor
ferroelectric 0.9~PbMg$_{1/3}$Nb$_{2/3}$O$_3$~-~0.1~PbTiO$_3$ near room temperature}% Force line breaks with \\

\author{A.S. Mischenko$^{1,}$\footnote{Electronic address: am507@cam.ac.uk}, Q. Zhang$^2$, R.W. Whatmore$^{2, 3}$,
and N.D. Mathur$^1$}

\affiliation{$^{1)}$Department of Materials Science, Pembroke
Street, University of Cambridge, CB2 3QZ, Cambridge, United Kingdom}

\affiliation{$^{2)}$Department of Materials, Cranfield University,
Beds MK43 0AL, United Kingdom}

\affiliation{$^{3)}$Tyndall National Institute, Lee Maltings,
Prospect Row, Cork, Ireland}

%\date{\today}% It is always \today, today,
             %  but any date may be explicitly specified

\begin{abstract}
We have recently observed a giant electrocaloric effect (12~K in
25~V) in 350~nm sol-gel PbZr$_{0.95}$Ti$_{0.05}$O$_3$ films near the
ferroelectric Curie temperature of 242$^\circ$C. Here we demonstrate
a giant electrocaloric effect (5~K in 25~V) in 260~nm sol-gel films
of the relaxor ferroelectric
0.9~PbMg$_{1/3}$Nb$_{2/3}$O$_3$~-~0.1~PbTiO$_3$ near the Curie
temperature of 60$^\circ$C. This reduction in operating temperature
widens the potential for applications in novel cooling systems.
\end{abstract}

%\pacs{Valid PACS appear here}% PACS, the Physics and Astronomy
                             % Classification Scheme.
%\keywords{Suggested keywords}%Use showkeys class option if keyword
                              %display desired
\maketitle

\section{\label{sec:Intro}Introduction}

We have recently demonstrated~\cite{OurScience} a giant
electrocaloric (EC) effect in thin film
PbZr$_{0.95}$Ti$_{0.05}$O$_3$ (PZT)~\cite{Sawaguchi}, with a peak
$\Delta T$~=~12~K at 226$^\circ$C. This high working temperature
could permit cooling applications in the automotive, aerospace or
food industries, but a lower working temperature would open up many
more possibilities, e.g. on-chip refrigeration. Lower temperatures
are also attractive because thin films fatigue less quickly and
possess higher breakdown fields. Here we demonstrate an EC effect in
0.9~PbMg$_{1/3}$Nb$_{2/3}$O$_3$~-~0.1~PbTiO$_3$
(0.9~PMN~-~0.1~PT)~\cite{Smolensky} films that peaks at the
significantly lower temperature of 75$^\circ$C.

Relaxor ferroelectrics could be attractive for cooling applications.
Indeed, many relaxor films such as 0.9~PMN~-~0.1~PT show pronounced
pyroelectricity~\cite{ChoiBhalla, DavisSetter, WhatmorePyro},
suggesting that the converse EC effect is strong. Moreover, phase
transitions in relaxors are broad so the range of operating
temperatures is wide. However, the potential for relaxor
ferroelectric films in cooling applications has not been considered.

\section{\label{sec:PMN_PT}PMN-PT thin films}

The PMN~-~PT family, based on the relaxor PMN~\cite{Smolensky}, has
already been proposed for many
applications~\cite{YeShirane,ChoiBhalla,ViehlandHystFreq,DavisSetter}.
The PMN-rich relaxors, and compositions near the morphotropic phase
boundary at 0.65~PMN~-~0.35~PT, are promising for capacitors due to
their large dielectric constant, and also high-strain
actuators/transducers and prototype microelectromechanical systems
due to their piezoelectric properties~\cite{YeShirane}.

Bulk 0.9~PMN~-~0.1~PT is a rhombohedral (pseudocubic) relaxor
ferroelectric at room temperature~\cite{YeShirane,ZekriaGlazer}. On
heating above $T_C\sim$~60$^\circ$C, this structure transforms to a
cubic paraelectric phase~\cite{Smolensky,YeShirane,ZekriaGlazer}.
Because the material is a relaxor, the corresponding peak in the
dielectric constant is frequency dependent~\cite{Smolensky}, and
broad due to microscopic inhomogenities~\cite{Smolensky, ChoiBhalla,
Jim}. The pyroelectric properties of PMN ceramics under high DC bias
fields up to 100~kV~cm$^{-1}$ have been studied for thermal IR
detector applications~\cite{WhatmorePyro}. As there is no data for
PMN-PT thin films at the temperatures and high electric fields of
interest, the EC effect could not be predicted from the literature.

\section{\label{sec:Experimental}Experimental}
0.9~PMN~-~0.1~PT sols were prepared from Sigma-Aldrich precursors. A
mixture of Pb(OAc)$_2$~$\cdot$~3H$_2$O and Mg(OAc)$_2$ was dissolved
in acetic acid and distilled at 100$^\circ$C for 30~minutes.
20\%~excess Pb and 5\%~excess Mg were added to compensate for losses
during sintering. Separately, acetic acid and 2-methoxyethanol were
added to a mixture of Nb(OEt)$_5$ and Ti(O$^n$Bu)$_4$ and the
resulting solution was stirred at room temperature for 30~minutes.
The Pb/Mg and Nb/Ti based solutions were mixed and stirred at room
temperature. Formamide was added to the final solution to prevent
cracking during sintering.

Sols were passed through a 0.2~$\mu$m filter for spin-coating at
3000~rpm for 30~s onto Pt(111)/Ti/SiO2/Si(100) substrates that had
been rinsed with acetone and propanol. Layers of $\sim$50~nm were
obtained by pre-firing in air on a hotplate at 350$^\circ$C for
30~s, and then further annealing in a tube furnace at 750$^\circ$C
for 3~minutes. This procedure was repeated five times to obtain
$\sim$260~nm films.

Film structure was determined by x-ray diffraction on a Philips
diffractometer using Cu~K$\alpha$  radiation. $\theta-2\theta$ scans
corresponded to a polycrystalline perovskite phase with no preferred
orientations. The amount of pyrochlore phase was~$\sim$8\%. Pt top
electrodes of diameter 0.2~mm were sputter deposited through a
mechanical mask, and the bottom Pt electrode was contacted with
silver dag at a substrate edge. The dielectric constant and loss
tangent were measured using a HP 4192A Impedance Analyser at 100 kHz
and 100 mV ac amplitude. Hysteresis measurements were carried out at
10 kHz using a Radiant Technologies Precision Premier workstation
and a low temperature~(80$^\circ$C to~-200$^\circ$C) probe station.
The temperature of the sample was controlled via feedback from a
thermocouple, accurate to~0.3$^\circ$C, in contact with the sample.

\section{Results and interpretation}

\begin{figure}
\includegraphics{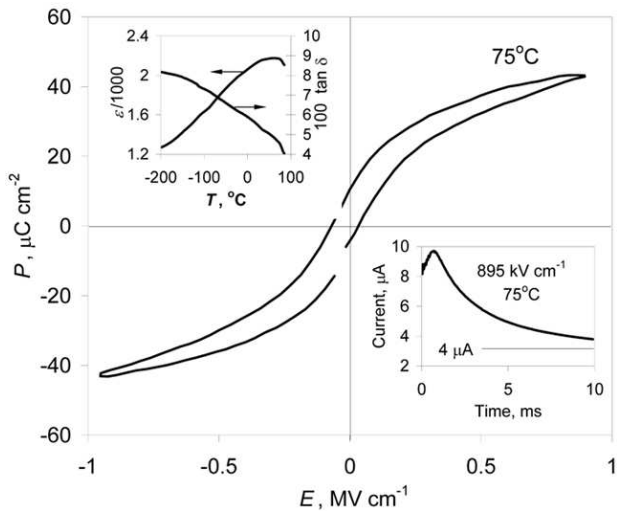}
\caption{\label{fig:Mischenko_Fig1}Electrical measurements of
0.9~PbMg$_{1/3}$Nb$_{2/3}$O$_3$~-~0.1~PbTiO$_3$ thin films.
Polarisation~$P$ versus applied electric field~$E$ at~10~kHz showing
evidence of ferroelectricity in the temperature range
studied~(80$^\circ$C to~-150$^\circ$C). Upper inset: the real part
of the effective dielectric constant measured at 100~kHz shows a
single broad peak at the bulk transition temperature
$T_C$~=~60$^\circ$C. Loss tangent tan~$\delta$~=~5\% at~60$^\circ$C.
Lower inset: leakage current measurements at the peak working
temperature $T_{Peak}$~=~75$^\circ$C.}
\end{figure}

Electrical hysteresis measurements were made roughly
every~$T$~=~10$^\circ$C in the range~80$^\circ$C to~-200$^\circ$C,
on cooling to minimise reductions in $P$ due to fatigue.
Fig.~\ref{fig:Mischenko_Fig1}~shows the expected~\cite{Smolensky}
ferroelectric~$P(E)$ at~75$^\circ$C, where the EC effect peaks as
shown later. Only one loop is shown because our measurements of
$P(E)$ in~80$^\circ$C to~-200$^\circ$C varied by just a few percent,
and are therefore visually similar. However, we show below that this
is sufficient to produce a giant EC effect. The real part of the
dielectric constant~$\varepsilon$ measured on cooling has a broad
peak at $T_C$~=~60$^\circ$C associated with the
ferroelectric-paraelectric transition
(Fig.~\ref{fig:Mischenko_Fig1}, upper inset). This broadness is
typical of relaxors due to microscopic inhomogeneity. It is also
typical of thin films due to interfacial strain, scalar
concentration gradients, or other forms of microscopic
variability~\cite{Smolensky, Jim}.

\begin{figure}
\includegraphics{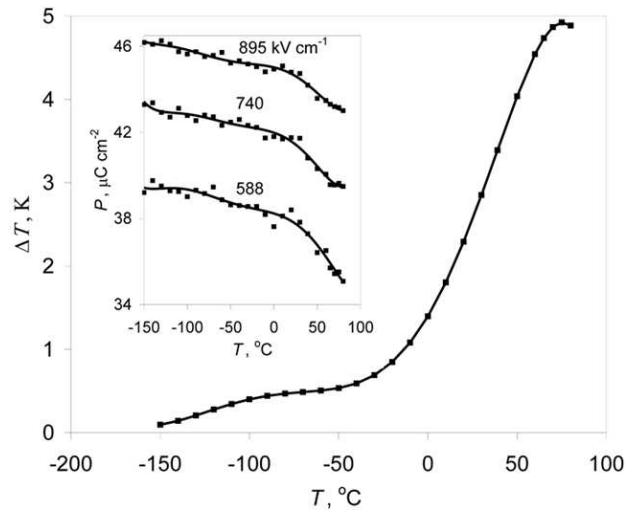}
\caption{\label{fig:Mischenko_Fig2}Electrocaloric temperature
changes. $\Delta T$ was calculated from equation (1) with applied
field $E$~=~895~kV~cm$^{-1}$. The peak value of~5~K occurs
at~$T_{Peak}$~=~75$^\circ$C. Inset: the temperature dependence of
polarisation~$P$ at selected~$E$. Data extracted for~$E$~$>$~0 from
the upper branches of~26 hysteresis loops measured at 10~kHz in
150$^\circ$C~$\leq$~$T$~$\leq$~80$^\circ$C. The EC effect is largest
when $\left| \frac{\partial P}{\partial T} \right|$ is maximised at
the broad paraelectric to ferroelectric transition. The lines
represent 6$^{th}$ order polynomial fits to the data.}
\end{figure}

Reversible adiabatic temperature changes~$\Delta T$ due to an
applied electric field~$E$, for a material of density~$\rho$ with
heat capacity~$C$ are given~\cite{TuttlePayne} by:

\begin{equation}
\Delta T =
-\frac{1}{\rho}\int_{0}^{E}{\frac{T}{C}\left(\frac{\partial
P}{\partial T}\right)_{E'}}dE',\label{eq:DT}
\end{equation}

assuming the Maxwell relation $\left( \frac{\partial P}{\partial T}
\right)_E = \left( \frac{\partial S}{\partial E} \right)_T$. Values
of $\frac{\partial P}{\partial T}$ were obtained from 6$^{th}$ order
polynomial fits to $P(T)$ data (Fig.~\ref{fig:Mischenko_Fig2},
inset). Fatigue may only reduce our values of $\left| \frac{\partial
P}{\partial T} \right|$ since the data were taken on cooling such
that~$P$ increased in successive hysteresis measurements. In the
temperature range of interest, the heat capacity
$C$~=~120~J~mol~$^{-1}$~K$^{-1}$ remains sensibly constant or even
decreases at low temperatures~\cite{GorevHeatCap}. We note that
assuming a constant value of~$C$ despite a $\sim$~50\%
peak~\cite{TuttlePayne} resulted in excellent agreement with direct
EC measurements of $\Delta T$ in bulk
Pb$_{0.99}$Nb$_{0.02}$(Zr$_{0.75}$Sn$_{0.20}$Ti$_{0.05}$)$_{0.98}$O$_3$~\cite{TuttlePayne}.
We assume the bulk value of
$\rho$~=~8.08~g~cm$^{-3}$~\cite{SatoDensity}.

EC temperature changes obtained with~(1) are presented in
Fig.~\ref{fig:Mischenko_Fig2}. The largest change at
$T_{Peak}$~=~75$^\circ$C (5~K in 25~V, i.e. 0.2~K~V$^{-1}$) exceeds
the previous best results at these temperatures obtained in bulk
PMN-PT at 14$^\circ$C (1~K in 160~V, i.e.
0.006~K~V$^{-1}$)~\cite{PMNPT_ECE_Bulk} and bulk
PbSc$_{0.5}$Ta$_{0.5}$O$_3$ at 20$^\circ$C (1.5~K in 1500~V, i.e.
0.001~K~V$^{-1}$)~\cite{Shebanov}.

Our peak EC temperature change of  $\Delta T$~=~5~K, determined with
$E$~=~895~kV~cm$^{-1}$, represents a peak energy change $C \Delta
T$~=~1.86~kJ~kg$^{-1}$. The corresponding hysteresis loss was~24\%
of this figure, as determined from the area of the~75$^\circ$C
hysteresis loop in~$E$~$>$~0. Hysteresis losses may be
reduced~\cite{ViehlandHystFreq} by (i)~reducing the measurement
frequency, (ii)~introducing chemical substituents, and (iii)~process
control to modify microstructure.

Leakage current was measured at~$T_{Peak}$~=~75$^\circ$C under our
maximum value of~$E$~=~895~kV~cm$^{-1}$
(Fig.~\ref{fig:Mischenko_Fig1}, lower inset). Reliable measurements
were not possible beyond 10~ms, at which time the leakage current
has fallen to~4~$\mu$A. However, the graph shows that transient
currents continue to fall at this measurement time.
Therefore~4~$\mu$A represents an upper bound on the steady-state
leakage current. A current of this magnitude generates a Joule
heating of 0.1~K over one quarter of a cycle. This is negligible
compared with the peak EC effect of~5~K.

We have demonstrated here a giant EC effect in the relaxor
ferroelectric~0.9~PMN~-~0.1~PT. The effect peaks at~75$^\circ$C,
which is nearer to room temperature than the giant peak found for
PZT at 226$^\circ$C~\cite{OurScience}. Reducing the PT content in
PMN-PT is known to reduce the ferroelectric transition temperature
to~$\sim$0$^\circ$C in pure PMN~\cite{Smolensky}. This suggests a
means by which to achieve significant EC effects at even lower
temperatures.

\begin{acknowledgments}
A.M. was supported by Churchill College, Cambridge, an honorary
Kapitza Scholarship from the Cambridge Overseas Trust, and an
Overseas Research Scholarship award from Universities UK. Cranfield
University gratefully acknowledges financial support from the UK
EPSRC under the Platform Grant GR/R92448/01. We thank the UK EPSRC
for additional funding, and J.F.~Scott, F.D.~Morrison, G.~Catalan
and P.~Zubko for discussions.
\end{acknowledgments}

\bibliography{MischenkoPMN_PT}% Produces the bibliography via BibTeX.

\begin{thebibliography}{10}

\bibitem{OurScience}
A.S. Mischenko, Q.~Zhang, J.F. Scott, R.W. Whatmore, and N.D. Mathur.
\newblock {\em Science}, 311:1270, 2006.

\bibitem{Sawaguchi}
E~Sawaguchi.
\newblock {\em J. Phys. Soc. Japan}, 8:615--629, 1953.

\bibitem{Smolensky}
G.A. Smolensky.
\newblock {\em J. Phys. Soc. Japan}, 28:26, 1970.

\bibitem{ChoiBhalla}
S.W. Choi, T.R. Shrout, S.J. Jang, and A.S. Bhalla.
\newblock {\em Ferroelectrics}, 100:29, 1989.

\bibitem{DavisSetter}
M.~Davis, D.~Damjanovic, and N.~Setter.
\newblock {\em J. Appl. Phys.}, 96:2811, 2004.

\bibitem{WhatmorePyro}
R.W. Whatmore, P.C. Osbond, and N.M. Shorrocks.
\newblock {\em Ferroelectrics}, 76:351, 1987.

\bibitem{YeShirane}
Z.G. Ye, B.~Noheda, M.~Dong, D.~Cox, and G.~Shirane.
\newblock {\em Phys. Rev. B.}, 64:184114, 2001.

\bibitem{ViehlandHystFreq}
D.~Viehland and J.F. Li.
\newblock {\em J. Appl. Phys.}, 89:1826, 2001.

\bibitem{ZekriaGlazer}
D.~Zekria, V.A. Shuvaeva, and A.M. Glazer.
\newblock {\em J. Phys.: Cond. Matt.}, 17:1593, 2005.

\bibitem{Jim}
M.M. Saad, P.~Baxter, R.M. Bowman, J.M. Gregg, F.D. Morrison, and J.F. Scott.
\newblock {\em J. Phys.: Cond. Matt.}, 16:L451, 2004.

\bibitem{TuttlePayne}
B.~A. Tuttle and D.A Payne.
\newblock {\em Ferroelectrics}, 37:603--606, 1981.

\bibitem{GorevHeatCap}
M.V. Gorev, I.N. Flerov, V.S. Bondarev, and Ph. Sciau.
\newblock {\em J. Exp. Theor. Phys. (JETP)}, 96:531, 2003.

\bibitem{SatoDensity}
Y.~Sato, H.~Kanai, and Y.~Yamashita.
\newblock {\em J. Am. Ceram. Soc.}, 79:261, 1996.

\bibitem{PMNPT_ECE_Bulk}
L~Shaobo and L~Yanqiu.
\newblock {\em Mat. Sci. and Eng.}, 113:46, 2004.

\bibitem{Shebanov}
L.A. Shebanov, E.H. Birks, K.J. Borman, and A.R. Sternberg.
\newblock {\em Ferroelectrics}, 94:305, 1989.

\end{thebibliography}

\end{document}